# Estimating the damage of railway carriage wheels equipped with disc brakes

Márton Krénusz[1], Róbert Sándor Kovács[1,2*], Miklós Krémer[3]

[1] Department of Energy Engineering, Faculty of Mechanical Engineering, Budapest University of Technology and Economics, H-1521 Budapest, P.O.B. 91, Hungary
[2] Department of Theoretical Physics, Wigner Research Centre for Physics, H-1525 Budapest, Hungary
[3] Knorr-Bremse Rail Systems Budapest; Department of Railway Vehicles and Vehicle System Analysis, Faculty of Transportation Engineering and Vehicle Engineering, Budapest University of Technology and Economics, H-1521 Budapest, P.O.B. 91, Hungary
* Corresponding author, e-mail: kovacsrobert@energia.bme.hu

**Abstract**

The aim of this paper is to estimate the damage of railway carriage wheels caused by lock-up braking. First, we investigate the typical characteristics of wheel-rail contact including the parameters of contact patch, longitudinal creep and coefficient of adhesion. Then we show the current requirements concerning the wheel-slide protection systems and determine the points which should be reviewed. To predict the effect of the sliding phenomenon, material properties are needed, as well. Afterwards, a simplified thermal simulation is built to estimate where martensite formation may occur.

## 1 Introduction

Railway is one of the most used transportation facilities nowadays. Like other fields, trains are getting faster and faster every day, as well. The increased speed results in a higher demand on the requirements of the wheel-slide protection system. However, phenomenon of sliding is still difficult to eliminate. Especially when braking is too intense, and the coefficient of adhesion is low, lock-up of the wheels may occur. This means that the wheels have a sliding velocity of v but no angular velocity. As the wheel is sliding on the rail, a significant heat is being generated between the contacting bodies that results in high temperatures on the surface and near the contact patch. This may even cause changes in the microstructure of the wheel material and bring about the formation of a brittle and hard martensite layer. Thus, cracking and spalling may appear on the tread, so leading to wheel damage in the long run [1].

A simplified thermal model was used to estimate where phase changes were expected in case of lock-up braking. To predict the consequences of this type of sliding, a research was required on wheel-rail contact, wheel-slide protection system and wheel material.

## 2 Wheel-rail contact

In fact, both contacting bodies undergo elastic deformation due to the axle load. If we assume the validity of Hertzian contact mechanics, this results in an ellipse-shaped contact patch, which can be divided into two segments: an adhesion and a slip region. In the slip area local pressure is not high enough to transmit tangential force without slippage. In case of lock-up braking, the slip region covers the whole contact patch.

Wheel-rail contact is usually described using the coefficient of adhesion – longitudinal creep function. Latter depicts the virtual sliding velocity caused by deformation of the wheel near the contact patch. Only approximations are available for the function mentioned above. One of these can be seen in Figure 1 [2,3], where the region of lock-up braking is assigned, as well. It is clearly visible, that the boundary of pure sliding is located at the minimum (or maximum in case of driving) of the graph. After that, a significant decrease can be observed.

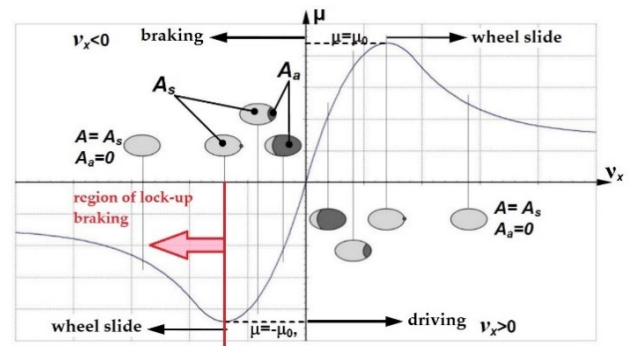

**Figure 1:** Graph of the coefficient of adhesion - creep function [2]

## 3 Wheel-slide protection

The wheel-slide protection (WSP) system is integrated in the basic brake equipment. Several requirements were set out regarding sliding. According to EN15595 and UIC541-05 standards, maximum sliding velocity can't exceed 30 km/h, while the duration of lock-up sliding can't be longer than 0.4 seconds. EN15595 restricts the amount of total energy per wheel, as well. This value is 26 kJ per contact point. In this paper we review whether these requirements can eliminate martensite formation completely [4].

## 4 Wheel material

Train wheels are mostly made of low alloy steel. For the simulation a steel with 0.5 % carbon content was used. As shown in the CCT curve in Figure 2, this material undergoes phase change at high temperatures. Above 735°C partial austenite formation can be observed, while over 780°C complete austenitization occurs. In case of rapid cooling, the martensite phase replaces the previously austenitized points [5].

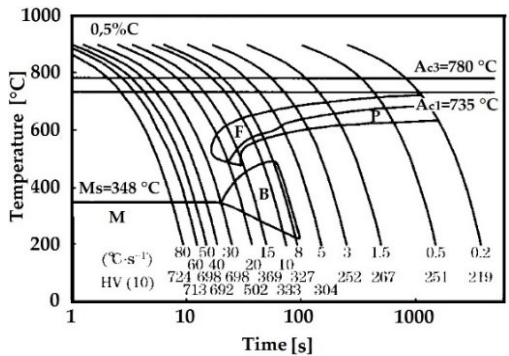

**Figure 2:** CCT curve of wheel material [5]

## 5 Simulation
### 5.1 Geometry

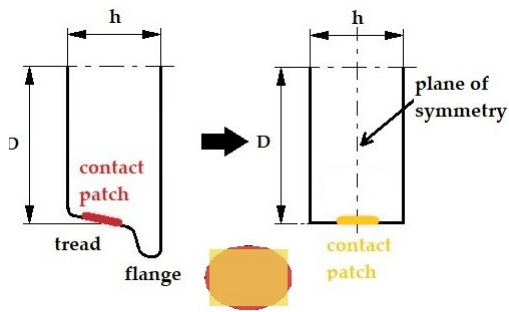

**Figure 3**: The simplified geometry and contact patch

For modelling a simplified geometry was built. As the size of the contact patch is small compared to the dimensions of the wheel, only the parts near the contact surface were taken into consideration. Thus, volumes far from this region including the flange were neglected. Special cone shape of the tread was also disregarded as it affects the results marginally. This led to a cylinder which could be cut in half according to Figure 3 due to the symmetric arrangement of the case. The contact ellipse was replaced with an equivalent rectangle which had the same area as the original surface [1].

### 5.2 Meshing

The meshing of the model was implemented using tetrahedral elements. The final meshed geometry can be seen in Figure 4. As near the contact patch large gradients were expected, denser mesh was applied in that region. The highest cell size was 1 mm while the smallest one we set to 0.15 mm, as Figure 4 shows, as well.

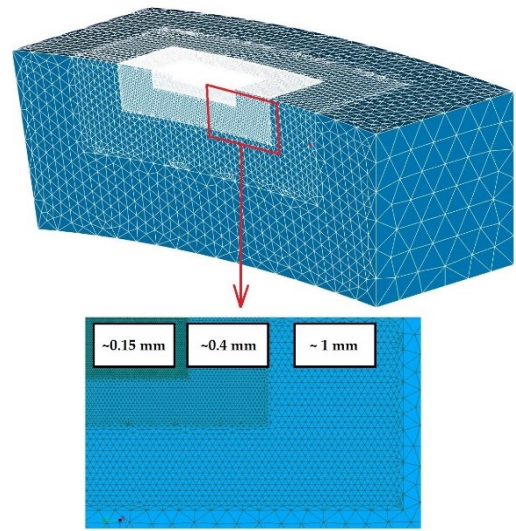

**Figure 4:** The meshed geometry

### 5.3 Thermal simulation

The simulation was carried out via open-source software programs. For preprocessing e.g. geometry modelling and meshing we used Salome. Elmer served as the solver for the thermal finite element equations. Results were evaluated in ParaView.

Input data were given according to the parameters of a real railway carriage as written in Table 1. Values of velocity and duration of sliding were set as written in the requirements of wheel slide protection (WSP) systems to check whether they were strict enough to prevent martensite formation. The chosen coefficient of adhesion is similar to the one used by Kennedy [6] to minimize errors coming from different adhesion parameters when applying the compensation factor mentioned in subsection 5.7.

| Mass of the vehicle | Number of axles | Sliding velocity | Duration of sliding |
|---|---|---|---|
| 50 000 kg | 4 | 30 km/h | 0.4 s |
| Nom. diam. of the wheel | Width of the wheel | Semi-minor axis of contact patch | Semi-major axis of contact patch |
| 840 mm | 140 mm | 8.82 mm | 11.04 mm |
| Coefficient of adhesion | Heat conductivity | Specific heat capacity | Density |
| 0.075 | 54 W/(mK) | 460 J/(kgK) | 7800 kg/m³ |

**Table 1** Input data for a general case of lock-up braking

### 5.4 Assumptions

During simulation several assumptions were set. For example, there was no heat loss towards the environment. We neglected the temperature dependence of material properties. Mechanical effects like stresses and strains were not taken into consideration. Coefficient of adhesion and sliding velocity were set constant. The traction was homogenous on the whole contact surface. A third layer e.g. some kind of fluid did not appear between contacting bodies.

### 5.5 Segments of simulation

The simulated process consists of three parts. Figure 5 shows the significant changes in the temperature of the point in the wheel near the contact patch. In segment I, the point goes through a rapid warm up due to the lock-up sliding condition. The wheel can be considered as a moving heat source from the rail's point of view. Therefore, the temperature of the contact patch tends to an equilibrium value in case of segment II, as the rate of conduction and cooling effect of rail are almost the same. The relationship between heat conductivity and vehicle speed characterizes this stage and determine, whether an equilibrium is reached. Afterwards, in segment III a significant cooling occurs when the wheel starts to rotate again [5].

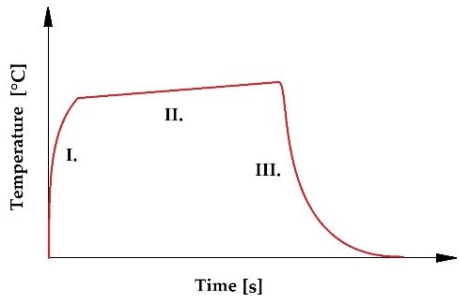

**Figure 5:** Three segments of simulation

### 5.6 Boundary Conditions

Boundary conditions were defined according to the geometrical simplifications as shown in Figure 6. On the contact patch we set heat flux during sliding and convective heat transfer in the cooling interval. The tread is in contact with the environment, so convective heat transfer occurs in this region. Zero heat flux was defined on the plane of symmetry. Constant temperature, 20 °C was applied on the rest of the boundaries. This is equal to the initial temperature value, as we assume, that these surfaces are so far from the contact patch that no temperature rise is expected there.

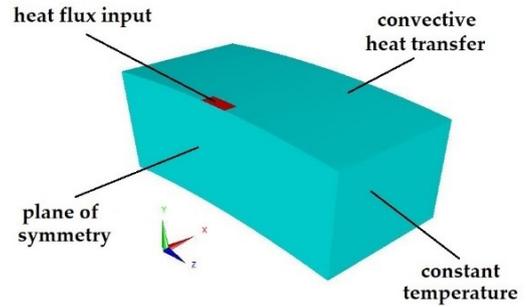

**Figure 6**: Boundary conditions

### 5.7 Input heat flux

If there is no heat dissipation, the amount of heat flux towards the wheel can be calculated from the power of friction force per area as shown in Equation 1. We assume that the normal load stems from the weight of the vehicle and it is distributed equally on each wheel. This leads to the following:

$$q = \frac{1}{A} \cdot \beta(x,t) \cdot \frac{m \cdot g}{2 \cdot n} \cdot \mu \cdot v, \qquad (1)$$

where A is the contact surface, m is the mass of the vehicle, n is the number of axles, μ is the coefficient of adhesion, v is the sliding velocity and β is the partition factor.

The partition factor β is due to the movement of wheel. As it is sliding, the hot front of the contact patch is permanently getting in contact with a cooler segment of the rail. Consequently, a compensational heat is going towards the rail thus reducing the amount of heat entering the wheel. This phenomenon can be described by the compensation factor which shows how many percent of the generated heat reaches the wheel. According to Kennedy [6], this parameter is a difficult function of time and the coordinates. Due to the limitations of Elmer the partition factor was approximated with linear functions as shown in Figure 7 [7].

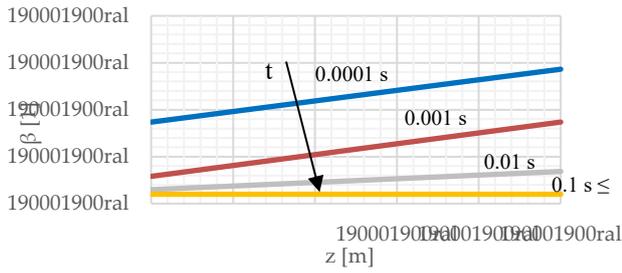

**Figure 7:** Time and spatial dependence of partition factor

The direct investigation of the rail was not taken into consideration as the temperature distribution in the rail is out of interest in this article. In addition to this, applying a two-body thermal model would have brought about several problems including for example the material and the shape of the rail, and the so-called third layer between contacting components. Also modelling would have become more difficult due to modified boundary conditions, meshing and coupling. This would have resulted in a need for further simplifications, thus leading to inaccuracies, as well. Based on this thought, it was more advantageous to include the effect of the rail via the empirical partition factor.

# 6 Results

## 6.1 Temperature distribution

The temperature distribution on the plane of symmetry can be seen on Figure 8. Temperature values from different depths are presented in Figure 9. It became clear based on the results that even the temperature of the points in depth of 0.5 mm reaches the region of complete austenitization.

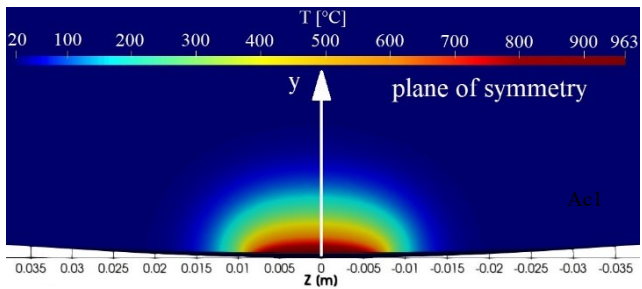

**Figure 8:** Temperature distribution on the plane of symmetry

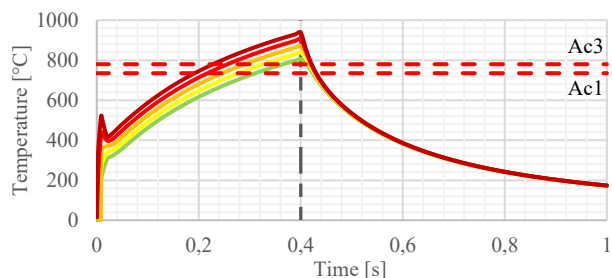

**Figure 9:** Temperature - time graph form the y axis

## 6.2 Cooling

To check whether martensite formation would occur, the cooldown intervals of the graphs above were placed on the logarithmic-scaled CCT curves.

It became clear that in every examined point the cooling was fast enough to produce martensite, as it can be seen in Figure 10.

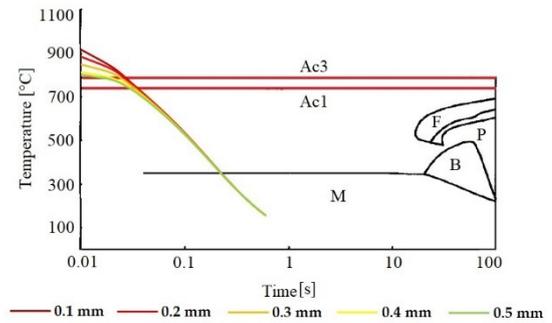

**Figure 10:** Cooling on the CCT curve

## 6.3 Region of possible martensite formation

We assumed that martensite would appear in each point where the temperature had reached the partial austenitic segment before. According to this, the region of possible martensite formation near the contact patch can be considered as a half ellipsoid. The x, y and z depths can be taken from the graphs in the Figures 12, 13, 14 based on the directions displayed in Figure 11.

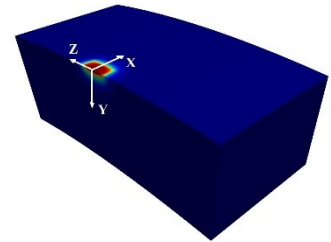

**Figure 11:** x,y,z direction displayed on the model

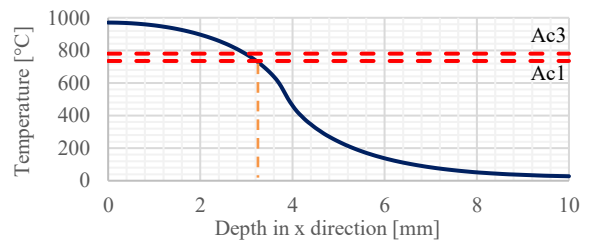

**Figure 12:** Temperature with respect to x depth

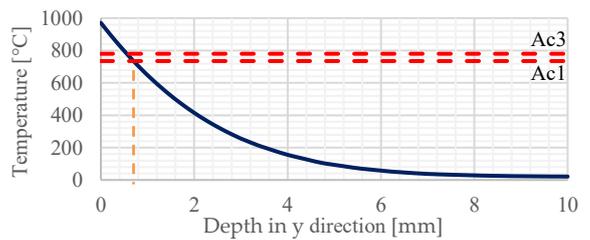

**Figure 13:** Temperature with respect to y depth

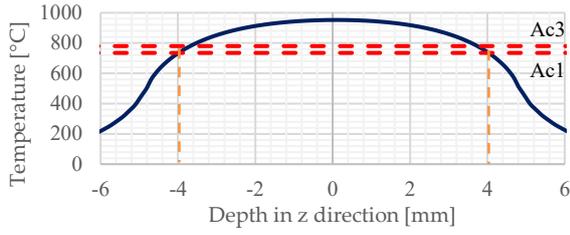

**Figure 14:** Temperature with respect to z depth

The results show that current requirements of wheel-slide protection system according to UIC541-05 do not prevent martensite formation near the tread. As mentioned before, EN15595 includes a limitation regarding total energy per wheel, as well. Equation 2 shows the method of calculating total energy:

$$E_{tot} = \frac{m \cdot g}{n} \cdot \mu \cdot v \cdot t, \qquad (2)$$

where t is the duration. In our case this value turned out to be according to Equation 3:

$$E_{tot} = \frac{50000 \cdot 9.81}{4} \cdot 0.075 \cdot \frac{30}{3.6} \cdot 0.4 \approx 31\ kJ, (3)$$

which also exceeds the required 26 kJ of total energy. This means, that in case of the parameters set to values in Table 1, martensite formation is expected. Further simulations were built to find out which values of sliding velocity and duration could result in a lower amount of generated total energy.

First, we ran two simulations with different durations: 0.3 s and 0.2 s. Other parameters including the sliding velocity were set the same as before. Then the total energy was calculated as shown in Equation 4 and 5:

$$E_{tot} = \frac{50\,000 \cdot 9.81}{4} \cdot 0.075 \cdot \frac{30}{3.6} \cdot 0.3 \approx 23\ kJ, \qquad (4)$$

$$E_{tot} = \frac{50\,000 \cdot 9.81}{4} \cdot 0.075 \cdot \frac{30}{3.6} \cdot 0.2 \approx 15\ kJ. \qquad (5)$$

In case of 0.3 s duration, the generated total energy is less than the prescribed value. However, it can be seen on Figure 15 and 16 that martensite formation still occurs if the lock-up braking lasts for 0.3 s.

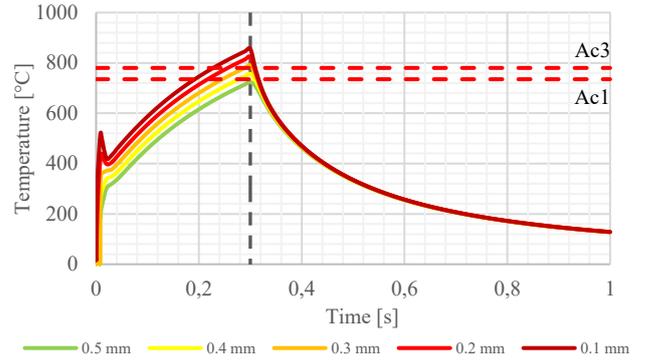

**Figure 15:** Simulation with 0.3 s duration

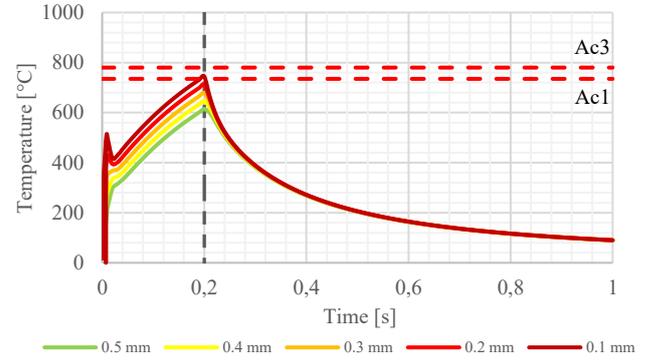

**Figure 16**: Simulation with 0.2 s duration

Afterwards, we ran 2 more simulations with different sliding velocities: 25 km/h and 20 km/h. The sliding was 0.4 s long to remove the effect of changes in duration. Equation 6 and 7 express the total energy:

$$E_{tot} = \frac{50\,000 \cdot 9.81}{4} \cdot 0.075 \cdot \frac{25}{3.6} \cdot 0.4 \approx 26\ kJ, \qquad (6)$$

$$E_{tot} = \frac{50\,000 \cdot 9.81}{4} \cdot 0.075 \cdot \frac{20}{3.6} \cdot 0.4 \approx 21\ kJ. \qquad (7)$$

If the speed was set to 25 km/h the total energy did not exceed 26 kJ, but according to Figure 17 and 18 appearance of martensite was still present.

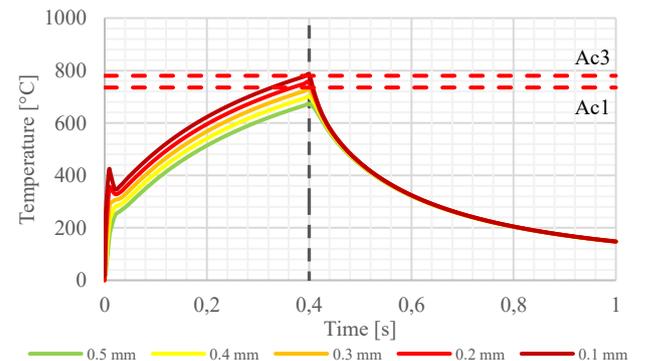

**Figure 17:** Simulation with 25 km/h velocity

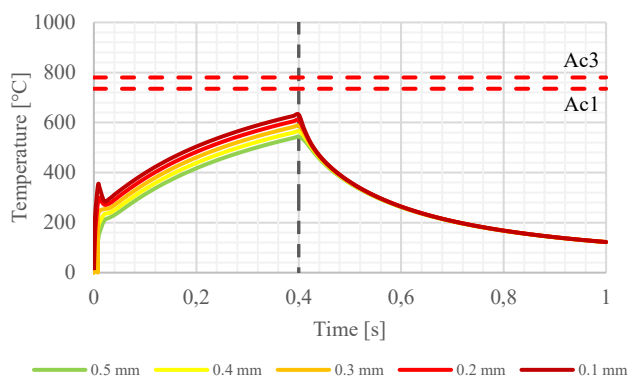

**Figure 18:** Simulation with 20 km/h velocity

## 7 Summary

| $V_{sliding}$ [km/h] | Duration [s] | Tot. Energy [kJ] |
|---|---|---|
| 30 | 0.4 | 31 |
| 30 | 0.3 | 23 |
| 30 | 0.2 | 15 |
| 25 | 0.4 | 26 |
| 20 | 0.4 | 21 |

**Table 2:** Summary of total energy values

It became clear that neither UIC541-05 nor EN15595 contains requirements that are strict enough to eliminate martensite formation completely as it can be seen in Table 2 which contains the effect of changes in sliding velocity and duration while other parameters are kept constant. In two cases (signed with light red background) the amount of total energy reaches or even exceeds the 26 kJ limit. Thus, even if the wheel slide protection system operates properly, a thin layer of martensite may appear near the contact patch because of the lock-up braking. This indicates, that from a thermal point of view, a controlled martensite formation is allowed according to the standards. To estimate the damage of the wheels, a further mechanical simulation and a fatigue test will be needed.

**Acknowledgement**

The research reported in this paper was supported by the grants of National Research, Development and Innovation Office—NKFIH FK 134277.
This paper was supported by the János Bolyai Research Scholarship of the Hungarian Academy of Sciences.